\documentclass[aps,prb,twocolumn,showpacs,floats]{revtex4-1}
\usepackage[dvips]{graphicx}
\usepackage{amsmath}
\usepackage{amssymb}
\usepackage{empheq}
\usepackage{dsfont}
\newcommand{\ua}{\uparrow}
\newcommand{\da}{\downarrow}
\newcommand{\ra}{\rangle}
\newcommand{\la}{\langle}
\newcommand{\ud}{\mathrm{d}}
\begin{document}

\title{\bf Ultrafast magnon-transistor at room temperature}

\author{Kevin A. van Hoogdalem and Daniel Loss}

\affiliation{Department   of  Physics,   University   of  Basel,
  Klingelbergstrasse      82,     CH-4056      Basel,     Switzerland}

\date{\today}

\begin{abstract}
We study sequential tunneling of magnetic excitations in nonitinerant systems through triangular molecular magnets. It is known that the quantum state of these molecular magnets can be controlled by application of an electric- or a magnetic field. Here, we use this fact to control the flow of a pure magnetization current through the molecular magnet by electric- or magnetic means. This allows us to design a system that behaves as a magnon-transistor. We show how to combine three magnon-transistors to form a NAND-gate, and give several possible realizations of the latter, one of which could function at room temperature using transistors with a 11 ns switching time.
\end{abstract}

\pacs{75.50.Xx, 75.30.Ds, 75.76.+j}
\maketitle
\section{Introduction}
In spintronic devices in insulating magnets, information about the logic state can be encoded in collective magnetic excitations, typically either spinons or magnons.~\cite{Mat81} Due to the nature of these carriers, power dissipation in such nonitinerant devices is anticipated to be much lower~\cite{Hal06,Tra08} than in their electronic counterparts, as well as in spintronics devices in semiconductors.~\cite{Wol01} As excess heating is a limiting factor in modern electronics, spintronics in insulating magnets is considered a contender to become the next computing paradigm.~\cite{Mei03,Hei04,Kos05,Imr06,Kaj10,Uch10,Kha11,Hoo11}

Since any classical algorithm can be implemented using a combination of transistors only, the design of this element in insulating magnets is a pivotal issue. Here, we theoretically show that that it is possible to make a transistor in which the logic state is encoded in purely magnetic excitations, and whose operation can be controlled by either a magnetic- or electric field. In our transistor, triangular molecular magnets~\cite{Gat07,Ses93,Tho96,Fri96,Ard07} take the role of gate, and we model the source and drain by spin reservoirs. We show that our transistor, which could operate at high clock speed at room temperature, can be used to implement the NAND-gate, one of the two existing universal gates for classical computation.

One of the earliest proposals for a spin-based logic device is the spin-field-effect transistor due to Datta and Das.~\cite{Dat90,Koo09} Other proposals include spin-field-effect transistors in non-ballistic systems~\cite{Sch03} and rings,~\cite{Aha11} a spin Hall effect transistor,~\cite{Wun10} an adiabatic spin transistor,~\cite{Bet12} and a bipolar magnetic junction transistor.~\cite{Fab03,Fab04,Ran10} However, all these proposals have in common the fact that they concern spintronics in semiconductors. In contrast, for the reasons explained above, we focus on spintronics in magnet insulators.

In our system, transport of magnetization occurs primarily by sequential tunneling of magnons (for ferromagnetic reservoirs) or spinons (for antiferromagnetic reservoirs) through a molecular magnet. We will show how it is possible to suppress or increase this sequential tunneling (and thereby switch between the insulating and conducting state of the transistor) by changing the internal state of the molecule through external fields, either electric or magnetic. Molecular magnets are especially suitable due to their chemical variety and controllability, as well as their relatively large size, which makes control of the state easier. For similar reasons, they have been proposed as good building blocks for novel spin-polarized-,~\cite{Bog08} as well as quantum computing devices.~\cite{Leu01,Leh07,Tro05}

This work is organized as follows: In Sec. \ref{sec:Sys} we introduce in more detail the previously mentioned system in which we will create our transistor. In Sec. \ref{sec:Rates} we calculate the tunneling rates of magnons/spinons through a triangular molecular magnet, and calculate the spin current through the molecular magnet from these rates. In Sec. \ref{sec:Tran} we show how controlling the state of a molecular magnet by electric- or magnetic fields allows us to design a transistor for either magnons or spinons. In Sec. \ref{sec:Exp} we focus on possible implementations of our transistor. Finally, we discuss certain constraints on our results in Sec. \ref{sec:Disc}.

\section{System}
\label{sec:Sys}
\begin{figure*}[t!]
\centering
\includegraphics[width=1.0\textwidth]{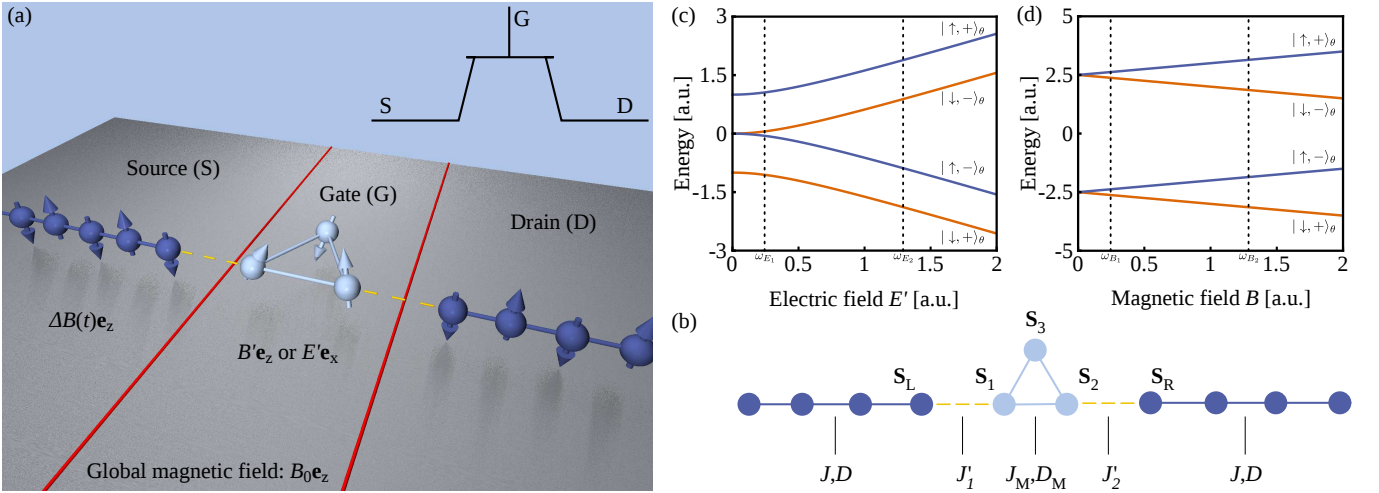}
\caption[]{(a)-(b) Pictorial representation of a single purely magnetic spin transistor, including Heisenberg parameters of the subsystems and the different magnetic- and electric fields. The field $B_0$ is applied to both spin reservoirs (here shown as 1D AF spin chains) and the molecular magnet, the magnetic field $\Delta B(t)$ is applied only to the left reservoir, and the fields $E'$ and $B'$ are applied only to the molecular magnet. The left reservoir acts as source-terminal of the transistor, the molecular magnet as gate, and the right reservoir as drain (see inset). (c)-(d) Excitation spectrum corresponding to the Hamiltonian in Eq. (\ref{eq:Hmol}). In (c) we choose $g_M\mu_B B = D_M = 1$, in (d) we put $D_M = 5$ and $E'=0$.}
\label{fig:Fig1}
\end{figure*}
The system we employ consists of a triangular molecular magnet such as \{Cu$_3$\} (see Ref. \onlinecite{Cho96}), weakly exchange-coupled to two identical spin reservoirs. Initially, we will assume that a single spin in each reservoir is coupled to a single spin located on a vertex of the triangular molecule, see Fig. \ref{fig:Fig1}(a)-(b). We will consider both one-dimensional (1D) and two-dimensional (2D) ferromagnetic (FM) spin reservoirs as well as 1D antiferromagnetic (AF) spin reservoirs. 

Both spin reservoirs, the molecular magnet, and the weak coupling between the subsystems are described by the isotropic Heisenberg Hamiltonian with Dzyaloshinkii-Moriya (DM) interaction
\begin{equation}
H =\sum_{\la ij\ra} J_{ij}{\bf s}_i \cdot {\bf s}_j + {\bf D}_{ij}\cdot\left({\bf s}_i \times {\bf s}_j\right).
\label{eq:HDM}
\end{equation}
 The exchange interaction $J_{ij}$ and DM vector $\textbf{D}_{ij} = D_{ij}\hat{\textbf{e}}_z$ are assumed to be constant for each subsystem. We assume that for the reservoirs $J_{ij} = J$ and $D_{ij} = D$, for the molecule $J_{ij} = J_M$ and $D_{ij} = D_M/\sqrt{3}$, and the coupling between the molecule and the left (right) reservoir is described by Eq. (\ref{eq:HDM}) with $J_{ij} = J'_{1(2)}$ and $D_{ij} = 0$ [see Fig. \ref{fig:Fig1}(b)]. We will assume that the $J'_i$'s set the smallest energy scale in the system, so that we can analyze tunneling processes using perturbation theory. In our model, a finite spin current is induced by application of a magnetic field $\Delta B$ to the left spin reservoir, which creates a non-equilibrium distribution of magnetic excitations. In reality, due to the finite lifetime of the magnetic excitations, a steady state spin current has to be generated using {\it e.g.} an AC magnet field difference, a static temperature difference,~\cite{Uch10} or spin pumping.~\cite{Kaj10,Tse02,Nak11}

When $D_M \ll J_M$, the low-energy subspace of the triangular magnetic molecule consists of a quadruplet with total spin 1/2, and the eigenstates of the Hamiltonian can be labeled as $|m_S,m_C\ra$ (see Refs. \onlinecite{Tri08,Tri10}). These states are eigenstates of the $z$ projections of two mutually commuting effective spin-1/2 degrees of freedom: the total-spin operator $\textbf{S}={\bf s}_1+{\bf s}_2+{\bf s}_3$ (with eigenvalues $m_S = \pm 1/2$) and the chirality operator $\textbf{C}$ (with eigenvalues $m_C = \pm 1$). The chirality operator can be defined by its $z$ component: $C^z = (4/\sqrt{3}){\bf s}_1 \cdot ({\bf s}_2 \times {\bf s}_3)$. An in-plane electric field $\textbf{E} = E^x\hat{\textbf{e}}_x+ E^y\hat{\textbf{e}}_y$ couples to the chirality through the spin-electric effect;~\cite{Tri08,Tri10,Isl10} the coupling of a magnetic field $\textbf{B}=B\hat{\textbf{e}}_z$ to the total spin is described by the Zeeman interaction. The Hamiltonian for the low-energy subspace is~\cite{Tri08,Tri10}
\begin{equation}
H_M = g_M \mu_B B S^z + d \mathbf{E}'\cdot \mathbf{C}^\parallel + D_M S^zC^z. 
\label{eq:Hmol}
\end{equation}
We split the magnetic field $\textbf{B}=B\hat{\textbf{e}}_z$ that is applied to the molecular magnet into two contributions, $B = B' + B_0$, where $B'$ is the magnitude of a local magnetic field that is applied only to the molecular magnet, and $B_0$ is the magnitude of a global magnetic field that is applied to the entire system [see Fig. \ref{fig:Fig1}(a)]. Furthermore, $d$ is the effective dipole moment of the molecule, $E'$ is the rotated electric field,~\cite{Tri08,Tri10} and ${\bf C}^\parallel = C^x \hat{\textbf{e}}_x + C^y \hat{\textbf{e}}_y$. The $g$-factor of the molecular magnet is denoted $g_M$.

Assuming that  $\textbf{E}' = E' \hat{\textbf{e}}_x$, we can rewrite Eq. (\ref{eq:Hmol}) in diagonal form as
\begin{equation}
H_M = g_M \mu_B B S_\theta^z + 2\sqrt{(D_M/2)^2+(dE')^2} S_\theta^z C_\theta^z.
\label{eq:Hmoldiag}
\end{equation}
Here, ${\bf S}_\theta$ (${\bf C}_\theta$) denotes the total-spin (chirality) operator with respect to the basis given by $|\ua, +\ra_\theta = \sin \theta |\ua,+\ra + \cos \theta |\ua,-\ra$, $|\ua,-\ra_\theta = -\cos \theta |\ua,+\ra + \sin \theta |\ua, -\ra$, and $|\da,+\ra_\theta$, $|\da,-\ra_\theta$ the same but with the total spin flipped and $\theta \to -\theta$. We defined $\tan \theta = [ \sqrt{(D_M/2)^2+(dE')^2}+D_M/2]/dE'$. The spectrum of the triangular molecular magnet is depicted in Fig. \ref{fig:Fig1}(c)-(d).

In the setup depicted in Fig. \ref{fig:Fig1}(a), tunneling of magnetic excitations between the left (right) spin reservoir and the molecular magnet is described by the isotropic Heisenberg exchange interaction between $\textbf{s}_{\textrm{L(R)}}$ and $\textbf{s}_{1(2)}$ [see Fig. \ref{fig:Fig1}(b)]. The corresponding Hamiltonian is given by $H_{\textrm{L}(\textrm{R})} = J_{1(2)}'{\bf s}_{\textrm{L(R)}}\cdot {\bf s}_{1(2)}$. Later, we will also consider the scenario in which a spin ${\bf s}_\textrm{M}$ in a reservoir is coupled to the third vertex of the triangular molecular magnet, ${\bf s}_3$. The Hamiltonian that describes this coupling is given by $H_\textrm{M} = J_3'{\bf s}_\textrm{M} \cdot {\bf s}_3$.

We want to find the effective Hamiltonian that describes the coupling $J_i'{\bf s}_j\cdot {\bf s}_i $ between a spin ${\bf s}_i$ on the molecular magnet and a spin ${\bf s}_j$ in a reservoir (hence $\{i,j\}=\{1,\textrm{L}\},\{2,\textrm{R}\},\{3,\textrm{M}\}$) within the low-energy subspace of the molecular magnet spanned by the basis $|m_S,m_C\ra$ defined above. In doing so, we neglect transitions to the higher-lying ${\bf S} = 3/2$ subspace of the molecule. This is allowed as long as we restrict ourselves to energies much smaller than $J_M$. By evaluating all relevant matrix elements $\la m_S,m_C|J_i'{\bf s}_j \cdot {\bf s}_i|m_S' ,m_C'\ra$, we find the effective Hamiltonian
\begin{equation}
H_j = {\bf s}_j \cdot \bar{\bar{J}}_i({\bf C}_\theta)\cdot {\bf S}_\theta + K_is_j^zC_\theta^z.
\label{eq:Htun}
\end{equation}
Here, $\bar{\bar{J}}_i({\bf C}_\theta)$ is an antisymmetric $3\times3$-matrix. We interpret the first term in Eq. (\ref{eq:Htun}) as effectively describing tunneling of spin excitations from the spin reservoir onto the total spin of the molecular magnet and \textit{vice versa}, with a tunneling strength that depends on the chirality state of the molecule. This term leads therefore to magnetization transport and is the one of interest to us. The second term does not induce hopping of magnetic excitations. We will show at the end of this section that its main effect for FM reservoirs is that of a static perturbation on the chirality state of the molecule, due to the static equilibrium magnetization $S\hat{{\bf e}}_z$ of a FM reservoir. In Sec. \ref{sec:AFres} we will show that $s_j^z$ is not the most relevant operator (in the renormalization-group sense) for AF reservoirs, so that we can ignore the second term for AF reservoirs.

The matrix $\bar{\bar{J}}_i({\bf C}_\theta)$ can be written generally as
\begin{equation}
\bar{\bar{J}}_i({\bf C}_\theta) = \frac{J_i'}{3} 
\left(\begin{array}{ccc}
A_i & B_i & 0 \\ -B_i & A_i & 0 \\ 0 & 0 & C_i
\end{array}\right).
\end{equation}
We find that $A_1 = -\cos (2\theta) - 2 C_\theta^x$, $B_1 = -\sin (2\theta) C_\theta^y$, and $C_1 = 1+2\cos (2\theta) C_\theta^x$. Furthermore, $A_2 = -\cos (2\theta) + C_\theta^x - \sqrt{3}\cos (2\theta)C_\theta^y$, $B_2 = -\sqrt{3}\sin(2\theta)-\sin(2\theta)C_\theta^y$, and $C_2 = 1+\sqrt{3}C_\theta^y - \cos (2\theta)C_\theta^x$. Lastly, we find $A_3 = -\cos(2\theta) +C_\theta^x + \sqrt{3}\cos(2\theta)C_\theta^y$, $B_3 = \sqrt{3}\sin(2\theta) -\sin(2\theta)C_\theta^y$, and $C_3 = 1-\sqrt{3}C_\theta^y-\cos(2\theta)C_\theta^x$. The different functions $K_i$ are given by $K_1 = -\sin(2\theta)$ and $K_2 = K_3 = \sin(2\theta)/2$.

We note that in the scenario where a single reservoir-spin ${\bf s}_{\textrm{L(R)}}$ is exchange-coupled to all three spins ${\bf s}_i$ of the molecular magnet with equal strength $J'$,  and we also put $E'=0$, the effective Hamiltonian is simply $J' {\bf s}_\textrm{L(R)}\cdot {\bf S}_\theta$. In this case, the tunneling of magnetic excitations no longer depends on the chirality of the molecule.

The first thing we see from Eq. (\ref{eq:Htun}) is that a static equilibrium magnetization $S\hat{{\bf e}}_z$ of a reservoir acts as a constant perturbation on the state of the molecular magnet through the relevant exchange interaction between reservoir and molecule. We will show that it is possible to make this perturbation trivial, or even beneficial to our purposes, in all cases under consideration. Additional dynamics of the systems is due to the behavior of magnetic excitations which exist on top of the equilibrium magnetization. We will study this dynamics next.

\section{Transition rates and spin current}
\label{sec:Rates}
To determine the spin current through the molecular magnet for the setup in Fig. \ref{fig:Fig1}(a), we use a master-equation approach. We assume that energy is conserved in all tunneling processes, and ignore higher-order effects. The transition rates from initial state $|i\ra$ to final state $|f\ra$ due to tunneling processes between the left (right) reservoir and the molecule are denoted by $R_{if}^{L(R)}$. Using Fermi's golden rule, we can calculate $R_{if}^{L(R)}$ to second order in $J'_{1(2)}$. We find
\begin{equation}
 R_{if}^L = \frac{1}{\hbar^2} \int_{-\infty}^\infty \ud \tau \la i|H'_{\textrm{L}}(\tau) |f\ra \la f | H'_{\textrm{L}}(0)| i \ra.
\label{eq:RLif}
\end{equation}
The rates $R_{if}^R$ are given by Eq. (\ref{eq:RLif}) with $H'_{\textrm{L}}$ replaced by $H'_{\textrm{R}}$. The apostrophe denotes an operator in the interaction representation with respect to the Hamiltonian of the uncoupled subsystems. The nontrivial part of the problem reduces then to finding correlation functions such as $\la s^+_\textrm{L}(\tau)s^-_\textrm{L}(0)\ra$ in the spin reservoirs, where $s^{+(-)}_\textrm{L}(\tau)$ denotes the spin raising (lowering) operator. In the next two sections we will find the relevant expressions for both FM and AF spin reservoirs.

Calculation of the spin current requires both the transition rates as well as the probabilities $P_i$ that the molecule is in the state $|i\ra$. We define the vector $\textbf{P} = \left(P_{\ua +}, P_{\ua -}, P_{\da +}, P_{\da -}\right)$. The time evolution of the occupation probability vector $\textbf{P}$ is then given by $\ud \textbf{P}/ \ud t = \hat{R}\textbf{P}$, where $\hat{R}$ is the 4x4 matrix that contains the appropriate transition rates. The steady state probabilities are contained in the kernel of $\hat{R}$, normalized such that $\sum_i P_i = 1$. Hence, $\textbf{P}$ is uniquely determined by the transition rates. The spin current $I_S$ is then defined as the net rate with which excitations leave the left reservoir
\begin{equation}
\begin{split}
 I_S & = \left(R_{\da + \ua +}^L + R_{\da + \ua -}^L\right)P_{\da +} + \left(R_{\da - \ua +}^L + R_{\da - \ua -}^L\right)P_{\da -} + \\
& - \left(R_{\ua + \da +}^L + R_{\ua + \da -}^L\right)P_{\ua +} - \left(R_{\ua - \da +}^L + R_{\ua - \da -}^L\right)P_{\ua -}.
\end{split}
\end{equation}
Next, we will calculate the relevant transition rates for the different types of reservoirs.

\subsection{FM reservoirs}
We first consider the simplest case of tunneling of magnons between a triangular molecular magnet and 1D FM reservoirs. We assume for simplicity that the FM reservoirs are described by the isotropic Heisenberg Hamiltonian, {\it i.e.} $D=0$. Using the Holstein-Primakoff transformation,\cite{Mat81} we can map a 1D FM system with spins $S \gg 1/2$ on a system of non-interacting bosonic particles (magnons). In the presence of a magnetic field $B_0\hat{{\bf e}}_z$, these magnons have a dispersion $\hbar\omega_q = 4|J|S\sin^2(qa/2) + g_R\mu_B B_0$. Here, $q$ is the wave vector of the magnons, $a$ is the lattice spacing of the reservoir, and $g_R$ is the $g$-factor of the reservoir. The Holstein-Primakoff transformation allows us to obtain the correlation functions that are required to find the transition rates by rewriting $\la s^+_\textrm{L}(\tau)s^-_\textrm{L}(0)\ra$ etc. in terms of the bosonic operators $a_q^\dagger,a_q$. We find the rates
\begin{equation}
\begin{split}
R^L_{\ua + \da +} & =  \left(\frac{J_1' \xi_+^L}{3\hbar}\right)^2 \frac{S}{2\pi} K_\textrm{FM}(\omega_B+\omega_E-\omega_{\Delta B}), \\
R^L_{\ua + \da -} & = \left(\frac{J_1' \eta_+^L}{3\hbar}\right)^2 \frac{S}{2\pi}  K_\textrm{FM}(\omega_B-\omega_{\Delta B}).
\label{eq:RFM}
\end{split}
\end{equation}
The energy scales are given by $\hbar \omega_B = g_M \mu_B B$, $\hbar \omega_{\Delta B} = g_R \mu_B \Delta B$, and $\hbar \omega_E = 2\sqrt{(D_M/2)^2+(dE')^2}$. Also, $\xi_+^L = \cos(2\theta)$ and $\eta_+^L =2-\sin(2 \theta)$. Furthermore, $K_\textrm{FM}(\omega) = \rho_\textrm{1D}(\omega_q)\left.\left[1+n_B(\omega_q)\right]\right|_{\omega_q=\omega}$, where  $\rho_\textrm{1D}(\omega_q) = a\left|\partial \omega_q/\partial q\right|^{-1}$ is proportional to the density of states (DOS) in the reservoir, and $n_B(\omega_q)$ is the Bose-Einstein distribution of the magnons. 

To get the other rates, we note that the only effect of inverting the total spin while keeping chirality unchanged (in the initial and final state simultaneously) is to replace $1+n_B(\omega_q) \to n_B(\omega_q)$ as well as $\xi_+^L \to \xi_-^L = \xi_+^L$ and $\eta_+^L \to \eta_-^L = 2+\sin(2\theta)$ in Eqs. (\ref{eq:RFM}); inverting both the chiralities while keeping the total spin constant changes $\omega_E \to -\omega_E$, $\xi_+^L \to \xi_-^L$, and  $\eta_+^L \to \eta_-^L$; finally, flipping both total spins and chiralities simultaneously requires us to replace $1+n_B(\omega_q) \to n_B(\omega_q)$ and $\omega_E \to -\omega_E$.

To obtain the rates for tunneling between the molecule and the right reservoir, we put $\omega_{\Delta B} = 0$ in Eqs. (\ref{eq:RFM}) and the derived rates. Furthermore, we replace $\xi_\pm^L \to \xi_\pm^R = \left|\cos(2\theta) + i\sqrt{3} \sin(2\theta)\right|$, $\eta_\pm^L \to \eta_\pm^R = 2\left|1\pm\sin(2\theta) +i\sqrt{3}\cos(2\theta)\right|$, and $J_1' \to J_2'$.

The transition rates due to a coupling $H_\textrm{M}$ can be calculated analogously, and we refrain from repeating those steps here. For the choosen setup, the resulting rates due to a coupling $H_\textrm{M}$ are the same as those due to a coupling $H_\textrm{R}$, except for the replacement $J_2' \to J'_3$. We also mention here that in the remainder of this work we will always chose our parameters (specifically $J_1',J_2'$, and possibly $J_3'$) in such a way that processes that only flip the chirality but not the total spin [such processes would be described for instance by a term $s_j^z C_\theta^x$ in Eq. (\ref{eq:Htun})] cannot occur. Hence, we can put these rates to zero.

To obtain the rates for a system with 2D FM reservoirs, we simply replace $\rho_\textrm{1D}(\omega_q)$ in Eqs. (\ref{eq:RFM}) by the 2D DOS, which for small $|{\bf q}|$ is given by $\rho_\textrm{2D}(\omega_\textbf{q})=\hbar/(4 S|J|)$.

\subsection{AF reservoirs}
\label{sec:AFres}
Next, we will derive the rates for tunneling of spinons between semi-infinite AF spin-1/2 chains and a triangular molecular magnet. In order to do so, we start by giving a description of the spin chains in terms of Luttinger liquid theory, which turns out to be a convenient framework for our purpose. For concreteness, we focus on the description of the left spin chain. Eq. (\ref{eq:HDM}) can then be mapped on the anisotropic Heisenberg Hamiltonian with anisotropy $\Delta = J/\sqrt{J^2+D^2}$ by performing a position-dependent rotation in spin space. After performing a Jordan-Wigner transformation, taking the continuum limit of the resulting fermionic Hamiltonian, and subsequent bosonization, the resulting Hamiltonian describing the left spin chain [for which $x \in (-\infty,0)$] becomes~\cite{Gia03}
\begin{equation}
H_L = \frac{\hbar}{2\pi}\int_{-\infty}^0\ud x \left[ uK\left(\partial_x\vartheta(x)\right)^2+\frac{u}{K}\left(\partial_x\varphi(x)\right)^2\right].
\label{eq:HLL_SI}
\end{equation}
The bosonic density field $\varphi(x)$ and its conjugate momentum field $\vartheta(x)$ satisfy $\left[\varphi(x),\partial_{x'}\vartheta(x')\right]=i\pi \delta(x-x')$. The sound velocity $u$ of the bosonic excitations as well as the interaction parameter $K$ can be determined from the parameters $J$ and $D$ of the spin chain using Bethe Ansatz results.~\cite{Hal80} At the isotropic point $K=1/2$, and $K=1$ describes the free model. To analyze the hopping between spin chain and molecular magnet, we will need the continuum form of the spin operators in the spin chain. These are given by~\cite{Egg92}
\begin{equation}
\begin{split}
s^-(x) & = \frac{e^{i\vartheta(x)}}{\sqrt{2\pi a}}(-1)^x\left[\cos\left(2\varphi(x)-2k_Fx\right)+1\right],
\\
s^z(x) & = -\frac{1}{\pi}\partial_x\varphi(x) + \frac{1}{\pi a} \cos\left[2\varphi(x)-2k_Fx\right].
\label{eq:Sop}
\end{split}
\end{equation}
Here, $a$ is the lattice spacing of the spin chain, and $k_F=\pi/2a+g_R\mu_B B_0 /u\hbar$ is the Fermi wave vector. For simplicity, we will ignore the small contribution proportional to $B_0$ in the expression for the Fermi wave vector. Since the spin chain is semi-infinite (with its last spin at $x=-a$), we require that the wave function vanishes at the origin. This constrains the density field to a constant value at the origin, such that $\cos \varphi(0) = 0$. To analyze the behavior of the spin fields at the origin, we introduce chiral fields $\varphi_{L/R}(x) = K\vartheta(x)\pm \varphi(x)$, which are related on the entire space by the constraint on the density field at the origin.~\cite{Egg92} This allows us to map Eq. (\ref{eq:HLL_SI}) on a quadratic Hamiltonian that only depends on $\varphi_L(x)$. Performing a renormalization group (RG) analysis on the spin operators near the boundary then yields that $s^z(0)$ is a marginal operator, and $s^{\pm}(0)$ scales as $1-1/(2K)$, so that it is relevant for systems with finite DM interaction. Since $s^z (0)$ is less relevant than $s^\pm(0)$, we will ignore the former in our perturbative analysis (assuming low enough energies). This allows us to ignore perturbations due to the interaction between the molecular magnet and the $s^z(0)$ terms in Eq. (\ref{eq:Htun}) for AF reservoirs. Since the RG flow is stopped either by temperature or by the relevant energy scale $E_M$ of the molecular magnet, this gives the constraint on the tunneling $J_i' \left[\textrm{max}(k_BT,E_M)/J\right]^{-1+1/(2K)} \ll J$ for our sequential tunneling approach to be valid.

In order to calculate the required transition rates, we need to calculate spin-spin correlation functions at the boundary of the spin chain. Since the density field is constant at the boundary, the sole relevant correlation function is that of the momentum field $\vartheta(t) \equiv \vartheta(0,t)$. At finite temperature $T$, it is given by
\begin{equation}
\la \left[\vartheta(t)-\vartheta(0)\right]^2\ra = \frac{2}{K} \ln \left[ \left(\frac{i\hbar \omega_C}{\pi \theta_0}\right) \sinh \left(\frac{\pi \theta_0\left[t-i\delta\right]}{\hbar}\right)\right].
\label{eq:theta_cor}
\end{equation}
Here,  $\theta_0 = k_B T$, and $\omega_C$ is the UV-cutoff of the theory. For this model it is approximated as $\omega_C \approx J/\hbar$. $\delta$ is a positive infinitesimal. The analysis of the right spin chains goes along the same lines, and we will refrain from repeating the steps here.

To calculate the transition rates for AF reservoirs we substitute Eqs. (\ref{eq:Sop}) in Eq. (\ref{eq:RLif}). Using the correlation function Eq. (\ref{eq:theta_cor}) and the fact that $\varphi(x)$ is constant at the boundary then gives the rates
\begin{equation}
\begin{split}
R^L_{\ua + \da +} & = \left(\frac{J_1' \zeta_+^L}{3\hbar\omega_C}\right)^2K_\textrm{AF}(\omega_B+\omega_E-\omega_{\Delta B}), \\
R^L_{\ua + \da -} & = \left(\frac{J_1' \nu_+^L}{3\hbar\omega_C}\right)^2K_\textrm{AF}(\omega_B-\omega_{\Delta B}).
\label{eq:RAF}
\end{split}
\end{equation}
Here, $\zeta_+^L = d_L\cos(2\theta)/2$, $\nu_+^L = d_L[1-\sin(2\theta)/2]$, and $d_L =\sqrt{2/\pi}$. We have ignored a small $k_F$-dependent contribution to $d_L$ here. The function $K_\textrm{AF}(\omega)$ describes the influence of the spin chain on the transition rate and is given by
\begin{equation}
\begin{split}
K_\textrm{AF}(\omega) & = \omega_C^2 \int_{-\infty}^\infty \ud \tau e^{i\omega\tau}e^{-\frac{1}{2}\la\left[\vartheta(\tau)-\vartheta(0)\right]^2\ra}\\
& = \omega_T\left(\frac{\omega_T}{\omega_C}\right)^{-2+\frac{1}{K}}e^{\pi\omega/\omega_T}\frac{\left|\Gamma(1/(2K)+i\omega/\omega_T)\right|^2}{\Gamma(1/K)}.
\end{split}
\end{equation}
Here, $\omega_T = 2\pi \theta_0/\hbar$.

To get the other rates, we note that the only effect of inverting the total spin while keeping chirality unchanged (in the initial and final state simultaneously) is to change the sign of $\omega_{B}, \omega_{\Delta B}$, and $\omega_E$, as well as to $\zeta_+^L \to \zeta _-^L = \zeta_+^L$ and $\nu_+^L \to \nu_-^L = d_L[1+\sin(2\theta)/2]$; inverting both the chiralities while keeping the total spin constant changes $\omega_E \to -\omega_E$, $\zeta_+^L \to \zeta_-^L$, and  $\nu_+^L \to \nu_-^L$; finally, flipping both total spins and chiralities simultaneously requires us to change the sign of $\omega_B$ and $\omega_{\Delta B}$ only.

To obtain the rates with the respect to the right spin chain, we put $\omega_{\Delta B} = 0$ in Eqs. (\ref{eq:RAF}) and the derived rates. Furthermore, we replace $\zeta_{\pm}^L \to \zeta_\pm^R = d_R\left|\cos(2\theta)+i\sqrt{3}\sin(2\theta)\right|/2$, $\nu_\pm^L\to\nu_\pm^R=d_R|1\pm \sin(2\theta)+i\sqrt{3}\cos(2\theta)|/2$, and $J_1'\to J_2'$. The constant $d_R = d_L$.

The transition rates due to a coupling $H_\textrm{M}$ can be calculated analogously, and we refrain from repeating those steps here. As in the FM case, the resulting rates are the same as those due to a coupling $H_\textrm{R}$ for the choosen setup, except for the replacement $J_2' \to J'_3$.

We note that processes that flip chirality but not the total spin are proportional to $s_j^z$. Since we have shown that this operator is less relevant than $s_j^\pm$, we can ignore such processes and put these rates to zero.

At this point, we have derived the transition rates for both AF and FM reservoirs. As we have seen, the main difference in the resulting rates is the replacement of the bosonic DOS and distribution function by correlation functions typical for Luttinger liquid models. We note here that, at low energies, the bosonic character of the magnons in the FM system yields larger spin currents than the fermionic spinons, which is extremely beneficial for the application we have in mind here.

\section{Transistor behavior}
\label{sec:Tran}
\begin{figure*}[t!]
\centering
\includegraphics[width=1.0\textwidth]{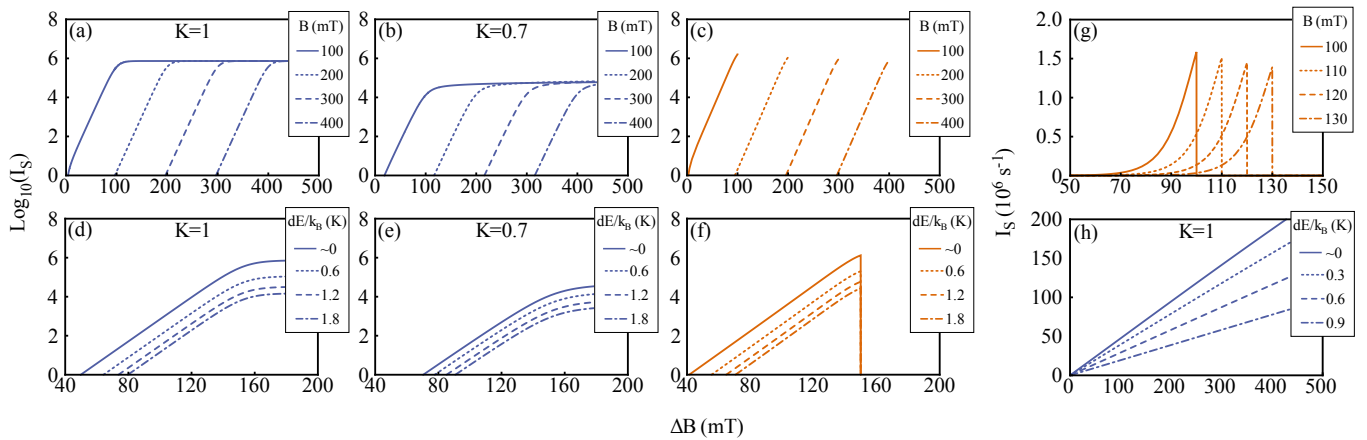}
\caption[]{(a)-(b) $\log_{10}(I_S)$ versus $\Delta B$ for different values of the magnetic field $B$, for 1D AF reservoirs with different Luttinger liquid parameters $K$. Due to the gapless nature of the spinons, we can always set $B=B_0$ and $B'=0$ when considering AF reservoirs. (c) $\log_{10}(I_S)$ versus $\Delta B$ for different values of the magnetic field $B$, for 1D FM reservoirs. Here, $B_0 = 1$ $\mu$T and $B'\approx B$. (d)-(e) $\log_{10}(I_S)$ versus $\Delta B$ for different values of $dE'$, for AF reservoirs with different Luttinger liquid parameters $K$. We assumed $D_M/k_B = 0.6$  K and $B_0 =150$ mT. (f) $\log_{10}(I_S)$ versus $\Delta B$ for different values of $dE'$, for the FM system. Here, $B_0 = 1$ $\mu$T and $B'=150$ mT. We assumed $D_M/k_B = 0.6$  K. (g) Illustration of the alternative switching mechanism for FM reservoirs. When the level splitting of the molecular magnet is smaller than the minimum energy of a magnon in the lead, the system is in the insulating phase. Again, $B_0 = 1$ $\mu$T. The plots (a)-(g) are for parameters $J/k_B = 100$ K, $T = 10$ mK, and $J'_1/k_B=J'_2/k_B = 0.05$ K (see text). For the FM plots, $S=1$. In (d)-(f) we have assumed that the left (right) reservoir is coupled to spin 2(3) in the molecular magnet with strength $J_{1(2)}'$. This setup is beneficial, since in this way both reservoirs decouple from the molecular magnet for $dE' \gg D_M$. (h) $I_S$ versus $\Delta B$ for different values of $dE'$ for the AF system, at an experimentally accessible temperature. Parameters are $J/k_B =100$ K, $J'/k_B = 2$ K, $D_M/k_B = 0.3$ K, $B = 75$ mT, and $T=1$ K.}
\label{fig:Fig2}
\end{figure*}
Next, we discuss two different ways in which our setup can be used as a logic switch whose working is controlled by an external magnetic field. This is one of two functionalities of a transistor, the other being that of amplification of a signal. We will briefly get back to this second functionality later in this section. For the description of the first mechanism by which our setup can be used as a logic switch, we assume that $k_B T \ll \hbar \omega _{\Delta B}, \hbar \omega_B \ll D_M$, so that we only need to take the states $|\ua,-\ra_\theta$ and $|\da,+\ra_\theta$ into account. We will assume $B_0 = 0$ in our explanation for simplicity. Referring back to Fig. \ref{fig:Fig1}(d), we see that spin transport through the molecule will be strongly suppressed for magnetic field differences such that $\omega_{\Delta B} \ll \omega_B$, since in this regime the vast majority of the excitations in the reservoirs lack the required energy to induce a spin-flip on the molecule; For $\omega_{\Delta B} \lesssim \omega_B$, transport increases rapidly with $\omega_{\Delta B}$. Hence, for magnetic field gradients $\omega_{\Delta B} \approx \omega_{B_1}$, our setup can be switched between the insulating- (for $\omega_B \approx \omega_{B_2}$) and conducting (for $\omega_B \approx \omega_{B_1}$) state. This is shown in Fig. \ref{fig:Fig2}(a)-(c).

The system with FM reservoirs offer an additional possibility to switch between the insulating and conducting state: When $\omega_{\Delta B} > \omega_B$, the minimum energy of the magnons in the reservoir exceeds the level splitting of the molecular magnet. In this case, the system is also insulating (neglecting higher order processes). This behavior has been indicated in Fig. \ref{fig:Fig2}(g). The use of this mechanism to switch between insulating and conducting states requires smaller magnetic fields compared to the previously discussed mechanism. Furthermore, this method does not require the assumptions on temperature put forward in the previous paragraph.

The second scheme may allow us to achieve amplification of an input signal in systems with FM reservoirs under certain conditions. We consider the magnetic field $B$ applied to the molecular magnet to be the input signal, and the magnetization of the drain to be the output signal. $\Delta B$ is assumed to be constant. It is important to remember that the bosonic distribution of magnons is peaked at low energies, {\it i.e.} $n_B(\omega_q) \gg 1$ at small $\omega_q$ and low temperatures. This translate into sizable currents at given (small) values of $\Delta B$ when the system is in the conducting phase. By switching $\omega_B$ between just below $\omega_{\Delta B}$ and just above $\omega_{\Delta B}$, we can then control this relatively large spin current by a small change in $B$. This can be viewed as a type of amplification. 

Lastly, we will discuss how the switching behavior can be controlled by an electric field. The mechanism is different from that for magnetic control, since an electric field does not influence the splitting between the two lowest states with opposite total spin. However, we note that when $k_B T, \hbar \omega _{\Delta B}, \hbar \omega_B\ll \hbar \omega_E$, transport occurs through transitions between states in the subspace spanned by $|\ua,-\ra_\theta$ and $|\da,+\ra_\theta$; transitions to the states $|\ua,+\ra_\theta$ and $|\da,-\ra_\theta$ are forbidden under these conditions since they cannot conserve energy. To illustrate the mechanism through which we can control the switching behavior by an electric field, we note that it follows from the expressions for the transition rates $R_{\ua-\da+}^R$ and $R_{\da+\ua-}^R$ in the previous sections that when $\theta \to \pi/4$, {\it i.e.} when $dE' \gg D_M$, the molecular magnet and the right spin reservoir are effectively decoupled in the low-energy subspace in the setup described there. This can be seen from the fact that the prefactors $\eta_-^R,\nu_-^R \to 0$ for $\theta \to \pi/4$.

Fig. \ref{fig:Fig2}(d)-(f) shows this switching behavior as a function of applied electric field. To determine the required strength of the electric field, we note that if we assume that the effective dipole moment lies between $d = (10^{-4} - 1) eR$ (see Ref. \onlinecite{Isl10}), where $R \approx 1$  nm is the bond length of the molecular magnet, then $dE'/k_B = 0.1$  K corresponds to an electric field $E' \sim (10^8 - 10^4)$ V m$^{-1}$.

In Fig. \ref{fig:Fig2} we assumed in plane Heisenberg exchange interaction between reservoirs and vertices for all AF reservoirs. This is motivated by the fact that the $s^z(0)$-operator in the AF reservoirs is irrelevant compared to the $s^\pm(0)$ operators. For FM reservoirs we assumed isotropic coupling $J'_1$ and $J'_2$. Additionally, we assumed that the third vertex of each molecular magnet is coupled to a separate reservoir by an Ising-like interaction with strength $J'_3=J'$. The reason behind this assumption is that in this way the sole effect of the equilibrium magnetization of the reservoirs on the Hamiltonian of the molecular magnets is to act as an effective magnetic field $J'S\hat{{\bf e}}_z$. This effective field for the parameters in Fig. \ref{fig:Fig2} (a)-(g) is on the order of 75 mT, and can take the role of $B'$. This reduces (or could even completely take over the role of) the required external magnetic field $B'$. We emphasize that the assumption of coupling to an additional third reservoir will not be needed for the experimental realizations of the magnon-transistor in the next section.

\section{Experimental realizations}
\label{sec:Exp}
\begin{figure*}[t!]
\centering
\includegraphics[width=1.0\textwidth]{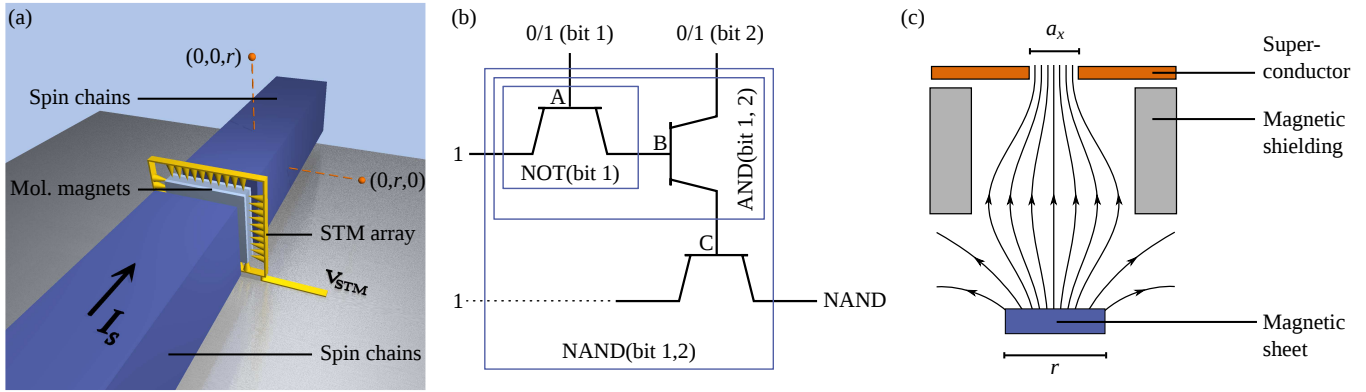}
\caption[]{(a) Proposed setup to measure the switching effect due to the spin-electric effect. The setup consists of a single-layer crystal of triangular magnets, of dimension 0.2 $\times$ 0.2 $\mu$m. The crystal contains $\sim 4 \cdot 10^4$ molecules, given a lattice constant of 1 nm. The crystal is weakly exchange-coupled to a bulk collection of AF 1D spin chains, such as is realized in SrCuO$_2$ (Ref. \onlinecite{Hlu10}) or Cs$_2$CoCl$_4$ (Ref. \onlinecite{Ken02}). Assuming the lattice constant is commensurate with that of the crystal, the setup contains $\sim 4 \cdot 10^4$ parallel transistors. The $\sim 10^3$ molecules at the edge can be accessed electrically by an array of STM tips. Using our previous estimate for $d$ and the values from the text, fields between $10^8-10^4$ V m$^{-1}$ are required to switch between insulating and conducting state. (b) Combining three transistors into a single NAND-gate. (c) Proposed setup to enhance the magnetic field due to the accumulated magnons. The role of the perforated superconductor is to increase the density of magnetic-field lines due to the magnetic field ${\bf B}_\textrm{mag}$ and thereby increase the magnetic field acting on the molecular magnet. The crystal of molecular magnets is placed directly above the hole in the superconductor.  We denote by $f$ the portion of the total flux directly on the surface of the magnet that can be enhanced into the area of the molecular magnets. $a_x$ and $a_y$ (not shown) are the dimensions of the hole of the superconductor, which should equal those of the crystal of molecular magnets.}
\label{fig:Fig3}
\end{figure*}
A single magnetic dipole moving with constant velocity $\textbf{v}$ gives rise to a magnetic field $\textbf{B}_\textrm{dip}(\textbf{r}) = \frac{\mu_0}{4\pi}\frac{g\mu_B}{r^3}\left[3\left(\hat{\textbf{e}}_z\cdot \hat{\textbf{e}}_r\right)\hat{\textbf{e}}_r-\hat{\textbf{e}}_z\right]$ in its rest frame, as well as an electric field $\textbf{E}_\textrm{dip}(\textbf{r}) = \textbf{v} \times \textbf{B}_\textrm{dip}(\textbf{r})$ in the laboratory frame. Conceivably, it is therefore possible to use the setup depicted in Fig. \ref{fig:Fig3}(a) to measure the switching behavior of a collection of spinon-transistors due to the spin-electric coupling at temperatures of $\sim 1$ K. This can be done by measuring the difference in voltage drop between points $(0,0,r)$ and $(0,r,0)$ in the insulating- [at $dE '\approx 2D_M$] and conducting (at $dE'\approx 0$) state. For the parameters in Fig. \ref{fig:Fig2}(h), and for $\Delta B \approx 200$ mT, the difference in spinon current
between the two states is $\sim 3\cdot 10^{10}$ spinons s$^{-1}$. For $r = 1$ $\mu$m, this leads to a difference in voltage drop of $\sim 10^{-13}$ V, which is within experimental reach.~\cite{Mik95} The strength of the required switching field $E'$ can be achieved near a STM-tip for molecules with reasonable $d$, see the caption of Fig. \ref{fig:Fig3}(a). This experiment would be interesting in its own right, since to our knowledge there have been no measurements of the spin-electric effect yet.

By combining three transistors as shown in Fig. \ref{fig:Fig3}(b), we can create a purely magnetic NAND-gate. The NAND-gate is a two-bit gate that gives a logical 0 as outcome if and only if both the input bits are 1, and yields a logical 1 otherwise. We will show how it may be possible to implement a NAND-gate consisting of magnon-transistors using readily available materials at $\sim 10$ K, and we will indicate the requirements for a working NAND-gate at room temperature.

Within a single transistor, the two FM spin reservoirs act respectively as source and drain. In our proposal, a finite non-equilibrium magnetization of the source (drain) encodes the logical state 1 of the source (drain); the logical state 0 has only the equilibrium magnetization present. A finite non-equilibrium magnetization of a reservoir is caused by having an excess number of magnons in that reservoir. The logic state at the gate is encoded in the strength of the local magnetic field, such that the transistor is insulating (conducting) if the logical state of the gate is 1 (0). Here, we propose to use the magnetic dipole field $\textbf{B}_\textrm{mag} = B_\textrm{mag} \hat{\textbf{e}}_z$ due to the excess $N$ accumulated magnons in the 1-state of the relevant terminal as gate-field at the points A-C in Fig. \ref{fig:Fig3}(b). 

We use the setup in which the left (right) reservoir in a given transistor is exchange-coupled to all three vertices of the triangular magnet of that transistor with equal strength $J_{1(2)}'$ (see Sec. \ref{sec:Sys}), and put $E'=0$. We assume that $J_1'>0$, so that the coupling between the left reservoir and the molecular magnet is antiferromagnetic; we set $J_2' < 0$, so that the coupling between the right reservoir and the molecular magnet is ferromagnetic. It is important to remember that both reservoirs are assumed to be ferromagnetic, but they are not necessarily identical. Therefore, they could be engineered in such a way that $J_1'$ and $J_2'$ have the properties stated above. In an experiment, $J_1'$ and $J_2'$ should be chosen such that the effects of the equilibrium magnetization of the reservoirs on the state of the molecular magnet (see end of Sec. \ref{sec:Sys}) approximately cancel each other, up to the required value of $B'$ in the conducting state of the transistor. This solves the issue of having to create relatively large local magnetic fields $B'$ in our proposal.

If the ordering of the energy levels of the molecule is such as depicted in Fig. \ref{fig:Fig1}(d), we only need to consider transitions between the states $|\da,+\ra_\theta$ and $|\ua, +\ra_\theta$. To switch between the insulating- and conducting state of a single transistor, we use the fact that the system is insulating for $\omega_E+\omega_{B_\textrm{mag}} > \omega_{\Delta B}$, and conducting for $\omega_E+\omega_{B_\textrm{mag}}< \omega_{\Delta B}$. Due to thermal fluctuations, the magnetic field $\textbf{B}_\textrm{mag}$ is not constant. These fluctuations limit the fidelity of our NAND-gate. We
characterize the fluctuations by the standard deviation $ \sigma = \la \left[\hat{n}-\la\hat{n}\ra \right]^2\ra^{1/2}$ of the number of magnons $n$ on the gate, and calculate $\sigma$ using the equilibrium distribution of the magnons using the grand canonical ensemble.

For the implementation at temperatures $T \sim 10$ K, we consider quasi-2D FM reservoirs of thickness $d$ with spin $S = 10$, $|J|/k_B = 5$  K, and lattice spacing $a = 1$  nm. These are the approximate values for yttrium iron garnet (YIG), a material that is often considered appropriate for applications in spintronics. For the source and drain, we consider a quasi-2D sample of dimensions 300 nm $\times$ 300  nm. For simplicity, we assume a single layer sample with $d = 1$ nm, and we put $J_1'/k_B=-J_2'/k_B = 1$ K in our calculation. As our gate, we use a single layer crystal of molecular magnets with $D_M /k_B = 0.26$ K (this corresponds to 200 mT), and $J_M/k_B \gg 10$ K. Lower $J_M$'s can be used when the the experiment is performed at a lower temperature, making the use of materials such as \{Cu$_3$\} (see Ref. \onlinecite{Cho96}) feasible. Assuming matching lattice constants, this setup contains 300 parallel single-molecule transistors. We estimate the field at the position of the molecular magnet [see Fig. \ref{fig:Fig3}(c)] due to the $N$ magnons as $\left|\textrm{B}_\textrm{mag}\right| \approx \frac{1}{2}\mu_0 N g\mu_B f /(d a_x a_y)$ (see caption of Fig. \ref{fig:Fig3}(c) for the definitions). Additionally, we use parameters $g_M = g_R = 2$, $f = 0.15$, $B_0 = 200$ mT, $B'=-160$ mT in the insulating phase and $B'=-140$ mT in the conducting phase, and $\Delta B = 50$ mT. In the conducting state, we find a magnon current exceeding $1.5\cdot 10^{10}$ s$^{-1}$, which amounts to a switching time of a transistor of $\sim 300$ ns. In our model, the fidelity of a single-NAND-gate then exceeds 99.9\%.

By using materials with an increased $g$-factor, we can create a NAND-gate that functions at room temperature, consisting of transistors with a $\sim 11$ ns switching time. We use the same setup as in the previous paragraph, but with parameters $S = 3$, $|J|/k_B = 500$ K, $a = 1$ nm for the 2D FM reservoir; $D_M/k_B = 2.6$ K (this corresponds to 200 mT for the $g$-factor under consideration), and $J_M/k_B \gg 300$ K for the molecular magnet; and $J_1'/k_B=-J_2'/k_B = 100$ K, $g_M = g_R = 20$, $f = 0.015$, $B_0 = 1$ T (easily achievable near the surface of a FM), $B'=-160$ mT in the insulating phase and $B'=-140$ mT in the conducting phase, and $\Delta B = 50$ mT. Evidently, the development of molecular magnets with an exchange interaction $J_M/k_B \gg 300$ K, needed in order to have stable molecular magnets at room temperature, will require a certain amount of experimental progress: typical values of currently existing molecular magnets are in the range of 1-10 K. However, there is nothing fundamental that forbids the existence of molecular magnets with larger exchange interaction. We find a magnon current exceeding $3.7 \cdot 10^{11}$ s$^{-1}$ in the conducting state, which corresponds to a 11 ns switching time. Reducing the magnon fluctuations on the gate can further reduce the switching time. As before, the fidelity of a single NAND-gate exceeds 99.9\%.

\section{Discussions}
\label{sec:Disc}
In this section, we will discuss several different requirements that have to be fulfilled for our perturbative calculations of the tunneling current through the molecular magnet to be valid. The first constraint on our calculations concerns the validity of our spin wave analysis of the FM spin chains; the number of magnons per site has to satisfy $\la a_i^\dagger a_i\ra \ll 2S$. In our calculations, the average number of magnons per site is typically ~ 0.05-0.13, so that non-interaction spin wave theory is valid. The AF theory is valid for energies much smaller than the exchange interaction $J$.

We checked the validity of our sequential tunneling approach in a self-consistent manner. For the AF reservoirs, the criterion is simply that the tunneling current is much smaller than the current in the ballistic system, that is $I_S(\Delta B) \ll g\mu_B \Delta B/h$.  For the FM reservoirs, we require that the broadening of the energy levels of the molecule is smaller than the unperturbed level splitting. In other words, all transition rates $R_{if}^{L/R}$ of the FM system satisfy
\begin{equation}
R_{if}^{L/R}/n_B(\omega_i-\omega_f) \ll |\epsilon^0_i-\epsilon^0_f|/\hbar,
\end{equation}
where $\epsilon^0_i$, $\epsilon^0_f$ are the unperturbed energies of the states $|i\ra$, $|f \ra$.

Another constraint is given by the fact that, near Breit-Wigner resonances, the current through the molecular magnet can be strongly increased due to coherent tunneling processes. This only holds at low temperatures, at higher temperatures the broadening of the thermal distribution destroys coherent tunneling, and the sequential tunneling approach is valid again. The minimal temperature $T$ for FM reservoirs is given by
\begin{equation}
R_{if}^{L/R}/n_B(\omega_i-\omega_f) \ll k_B T/\hbar.
\end{equation}
All our calculations are at high enough temperature for the sequential tunneling approach to be valid for the FM system.

We note that relaxation of the state of the molecular magnet can be neglected as long as the coupling strength between the reservoirs and the molecular magnet exceeds the coupling between the molecular magnet and hyperfine- and phonon baths.
\section{Conclusions}
Using a sequential tunneling approach, we have studied transport of magnons and spinons through a triangular molecular magnet which is weakly coupled to two spin reservoirs. We have shown that, by changing the state of the molecular magnet through application of an electric- or magnetic field, we can control the magnitude of the spin current through the molecular magnet. We used this fact to propose a magnon-transistor, whose operation can be controlled by an electric- or magnetic field. We have shown for which parameters our transistor could operate at room temperature with a ~11 ns switching time. We have shown how several magnon-transistors can be combined to create a NAND-gate.

\section{Acknowledgements}
This work has been supported by the Swiss NSF, the NCCR Nanoscience Basel, and the FP7-ICT project "ELFOS".

\end{document}